\documentclass[preprint,12pt,3p]{elsarticle}

\usepackage{moreverb,url}

\usepackage{subfigure}
\usepackage{graphicx}   
\usepackage{amsfonts}
\usepackage[]{algorithm2e}
\usepackage{graphicx}
\usepackage{subfigure} 
\usepackage{amssymb}

\journal{Ad Hoc Networks}

\begin{document}

\begin{frontmatter}

\title{Priority and collision avoidance system for traffic lights}

\author{Iv\'an Santos-Gonz\'alez, Pino Caballero-Gil, Alexandra Rivero-Garc\'ia and C\'andido Caballero-Gil}
\address{ \{ jsantos, pcaballe, ariverog, ccabgil  \} @ull.edu.es}
\address{Departamento de Ingenier\'ia Inform\'atica y de Sistemas\\ Universidad de La Laguna\\ Tenerife, Spain}

\begin{abstract}
In this paper,  a collision avoidance system is presented  to detect red light running and  warn nearby vehicles and pedestrians  in real time in order to prevent possible accidents. No complex infrastructure-based solution such as those based on radars or cameras is here required. Instead, a new solution based on smartphones carried by drivers and pedestrians is proposed so that it is the device inside the vehicle  violating a traffic light, the one that self-reports the offence in order to generate alerts  and warn  nearby vehicles and pedestrians  to prevent accidents. The proposal could also be used by road authorities to collect data on traffic lights that are most frequently violated  in order to define an action plan to investigate causes and look for  solutions. It includes a classifier for learning and estimating driver behaviour  based on collected data, which is used to predict whether he/she is about to run a red light or detect whether that has already happened. In the first case, the system broadcasts warnings directly to close vehicles  and pedestrians through Wi-Fi, while in the second case, the proposal warns vehicles and pedestrians in the neighbourhood through a server.  
The solution also includes a  prioritization system based on changing traffic lights at intersections according to the needs and characteristics of the traffic at all times, giving the top priority to emergency vehicles.
Furthermore, the proposal  involves the use of cryptographic schemes to protect authenticity and integrity of messages sent from traffic lights, smartphones and servers, and privacy and anonymity to promote the use of the system.  A beta version with some parts of the proposal has been implemented and the obtained results are promising.
\end{abstract}

\begin{keyword}
Red-light running \sep security \sep 
Wireless technologies \sep Security \sep 
Driving behaviour \sep  Wi-Fi
\end{keyword}

\end{frontmatter}

\section{Introduction}

Different road traffic problems have emerged around the world as a result of population growth, especially in large urbanized areas and dense populated areas.
Before autonomous cars are deployed and widespread, road safety depends mainly on drivers' decisions, and these depend only on what drivers see and hear.
This paper proposes a new tool based on interactive and cooperative driving that involves an effective traffic control system to provide additional data to drivers, which can be very useful for their decision-making.

The so-called Intelligent Transport System (ITS) is a set of technological solutions designed to optimize different modes of transport.
Its main goals are to improve  passenger comfort, increase travel safety,  mitigate traffic congestion and reduce fuel consumption. In order to achieve this,  vehicles and infrastructure must cooperate. Since  
cooperation is only possible if the involved entities can communicate,  wireless communications are required. Besides, ITS is based on different information technologies, sensors and the Internet. 

Vehicular Ad-hoc NETworks (VANETs) are a key part of the ITS, in which  information is exchanged  among vehicles and/or a communication infrastructure.  
Thus, it is assumed that  each vehicle has an information transmitter commonly known as On-Board Unit (OBU). 
As for  the communication infrastructure, it can be implemented in various ways. For instance, the infrastructure can be deployed  along the road as communication points commonly called Road Side Units (RSUs). Communication between OBUs is called Vehicle-TO-Vehicle (V2V), while communication between OBUs and RSUs is called Vehicle-TO-Infrastructure (V2I) and Infrastructure-TO-Vehicle (I2V). Finally, communication between OBUs and pedestrians is called Vehicle-TO-Pedestrian (V2P). Both the European and the non-European profiles for ITS, ITS-G5 and WAVE, are based on IEEE 802.11p to allow  V2V, V2I,  I2V  and V2P communications. 
Among the most adverse traffic circumstances, the problem known as red light running is one of the worst because it generates  a large number of accidents \cite{Retting}. In particular, it is the main cause of urban accidents in the United States  \cite{usareport}. 
In addition,  the Insurance Institute for Highway Safety (IIHS) \cite{IIHS} reports that half of those  killed in red light running crashes are not  offenders, but  pedestrians and other drivers  hit by red light runners.

The basic idea of this paper is to present a  priority and collision avoidance system for intersections, controlled by VANET elements  taking advantages of V2V and V2I communications. 
On the one hand, for the priority control we use real time information about the traffic. Currently, control over the traffic lights is static, that is, traffic lights change depending on time ranges that are programmed, but they do not take into account the state of roads in real time.
Thus,  traditional traffic lights do not consider any distinction between an emergency vehicles and other vehicles.
On the other hand, in order to address the problem of red light running,  different  solutions  have been proposed in the bibliography, such as new traffic signal mechanisms or red-light speed cameras to detect offenders. 
These solutions have proven to be effective in some cases, but too expensive to be widespread.

In this work, only V2I communications  are used to know  the location of vehicles all the time.
The system is designed to be used through common tools such as  smartphones so that the VANET is created with mobile devices.
Every driver uses a mobile  application to send its location data to a server through Internet. 
The developed application is a solution that allows knowing when someone has run a red light  with the aim of increasing road safety.
On the one hand, a collision avoidance system is presented to detect red light running and  warn nearby vehicles and pedestrians  in real time in order to try to prevent possible accidents.
The idea of increasing road safety is the main reason for having this application installed. However, in the future,  traffic authorities might force all drivers to have this application and  it could even be installed directly in smart vehicles.
On the other hand, the solution presents a new approach to a prioritization system based on changing traffic lights at intersections according to the needs and characteristics of traffic in each moment, giving the top priority to emergency vehicles.
In this case, the server acts as the controller that calculates the priorities so that it sends the right colour combination to the smart traffic light.

This paper is organized as follows. 
Section 2 provides   some background on related work. Section 3 briefly introduces the main ideas  of how the system works. In Section 4, specific details of the user application are presented. 
Section 5 explains the proposed theoretical solution  to protect the security of the scheme. 
Section 6 provides a short explanation of the implementation of the proposal and some data regarding  its performance. 
Finally, some conclusions and open problems close the paper in Section 7.

\section{Related Works}

The need to improve  road traffic management is evident throughout the world \cite{world}. 
Governments are concerned about the increasing number of vehicles on the road  and  traffic-related deaths.
For this reason, they are trying to improve traffic safety by exploring the potential of  ITS through numerous research projects  \cite{bin}. 

Early studies related to the specific problem of red light running focus on studying the  statistics of different characteristics and circumstances of  offenders, such as age, speed, time of  day or type of intersection \cite{Iswanjono} \cite{Martinez}  \cite{Porter} \cite{RettingUlmer} \cite{Romano}. 
Many camera-based red light running detection systems are controversial, and there is relatively little published literature on the used methodologies. The recent study \cite{Lavrenz} proposes a system that combines high-resolution signal controller data with conventional stop bar loop detection to identify vehicles that enter the intersection after the start of red, when many of the most serious red light running crashes occur. 
The paper \cite{Santiago} presents a software-based methodology that monitors the underlying dataset of existing radar-based vehicle detection infrastructure to continuously monitor red light running  incidents. In particular,  the trajectories of vehicles approaching an intersection were logged and combined with signal status information, so that by analysing the position of vehicles and signal status, red light running incidents were detected. 
None of the aforementioned studies proposes a method to predict red light running, nor present solutions to avoid possible collisions, which are the main goals of this work.

A prediction method based on a decision tree constructed with probabilistic models applied to the input obtained from a camera installed at the intersection  is  presented in \cite{Elmitiny} to provide a stop or go advice to   drivers approaching the intersection. 
Apart from the cost of camera-based solutions, compared with that work, the prediction method included in the first part of the proposal is more flexible  because it is based on machine learning algorithms. 
The recent paper \cite{Llau} proposes an accident prediction model based on the negative binomial distribution, built from motor vehicle crash data obtained by matching intersections with red light camera using geometric variables.

Machine learning is also used to predict red light running in \cite{Aoude}, based on a large set of real-world data extracted from radar sensors installed at an intersection.  A similar study based on  radar data   was performed in \cite{Zaheri}  using a decision tree. The paper
\cite{Li} applied artificial neural networks to predict red light running using simulated radar data. 
All  such  radar-based systems suffer from the so-called obstruction problem and can be very costly. The smartphone-based system presented in this paper does not have those limitations and drawbacks.
As for   countermeasures already  implemented in practice,  different  solutions  have been tested to try to reduce red light running. For instance, the work \cite{Rettingetal} includes several proposals, such as the replacement of signalized intersections by roundabouts, a more  adequate yellow signal time, and/or the addition of a brief phase with all signals in red.  
However, the  obtained results  show that none of these measures eliminates  the need for novel solutions to the problem of red light running.

One of the innovative ITS applications  to provide a solution to the problem of red light running  is the so-called intelligent traffic light, which can be self-controllable to maximize flow on the route and/or responsive to pedestrians needs so that, for instance, green light   changes  the duration for blind pedestrians. Besides, intelligent traffic lights can  be also used to punish  users who run red lights \cite{De Charette}  \cite{choi}.  
This solution, called red light camera, has been operating for several years in many different regions around the world \cite{red}. These traffic enforcement cameras  automatically capture an image of any vehicle  entering an intersection after running a red traffic light and send  this photo to the traffic signal control system so that the photo can be used as  evidence  assisting authorities in their enforcement of traffic laws. 
Generally, the camera is triggered when a vehicle enters the intersection (passes the stop-bar) after the traffic signal has turned red.
Different studies show  that  red light  cameras actually tend  to increase accidents  and are almost always boosted by monetary incentives. Another big problem is related to  privacy.
In the bibliography we can find different proposals. For instance, the works \cite{Wang} \cite{ZhangZhou} \cite{ZhangWang} propose  a collision avoidance system that uses a   distribution model to predict   red light running from   images obtained with a camera installed at the intersection, and I2V communications to warn nearby vehicles. In this case, the    problems related to the use of cameras mentioned above add up  to the current lack of I2V communications. 
The work  \cite{gradinescu}  presents  an adaptive traffic light system based on wireless V2V communication and fixed controller nodes deployed at intersections that determine  optimum values for the traffic lights phases. This system is based on short-range wireless V2V communications, which are not yet available. 

Another work \cite{trafficl} proposes the use of RFID for dynamic traffic light sequences to avoid problems that usually arise with systems that use image processing. RFID technology is applied to a multi-vehicle, multi-lane and multi-road junction area to provide an efficient time management scheme.  The conclusion is that the system can emulate the judgment of a traffic police officer on duty. The system  described here does not use RFID, and its main goal is not related to traffic management but to accident avoidance.

Google \cite{fairfield} presents several methods to automatically map  the three-dimensional positions of traffic lights and robustly detect traffic light status from the camera on board the vehicles. This work also encodes semantics or tags to indicate traffic lights that are difficult to detect in images. However, they used  these methods to map more than four thousand traffic lights, and to perform on board traffic light detection for thousands of drivers through intersections, and the test revealed problems due to lens flares or heavy rain.

A modern traffic light for six roads and four junctions has been implemented by programming in an integrated circuit  microcontroller \cite{kham}. Compared with the present paper, the focus there was on the problem of congestion   caused by long time delays of red traffic lights, while the main goal here is the reduction of traffic accidents caused by red light running.

Intersection safety has been addressed using a combination of different ITS technologies  through an initiative in   the United States called  Cooperative Intersection Collision Avoidance Systems  (CICAS)   \cite{CICAS} \cite{Goodwin}. According to its definition, the system allows providing real-time warnings on both   the vehicle and   the infrastructure. 
One of the safety applications   developed in CICAS is a Violation Warning System that allows the infrastructure to send status information of a traffic light to approaching vehicles so that, based on this information and  vehicle GPS, the system estimates the risk that the vehicle will violate the traffic light, and if this risk is high enough, it provides a warning to the driver. Although part of the goals and operation of this proposal coincides with those of CICAS, the ITS
technologies used there include  vehicle positioning, roadside sensors, and bidirectional wireless communication based on   DSRC technology, while the system proposed here has fewer requirements  and is much cheaper.

In the recent paper \cite{Fu}, a Smart In-Vehicle Decision Support System supported by V2I communications is presented to help making better stop/go decisions in the indecision zone as a vehicle is approaching a signalized intersection. The decision support models are realized so that each decision rule is responsible to specific situations for making right decisions even without complete information. 

Prioritization at intersections has been a problem to solve since the beginning of our roads.
Through the years different solutions have been proposed to this problem, but with time the increase in the number of vehicles has made conflicts worse. 
The proposed solutions should take less time than the static traditional decision-making mechanisms, always with the right decision. 
This problem has been approached from different ways, such as fuzzy logic, neural networks and expert systems to control intersections.

Several fuzzy systems have been tested to know when the light has to be switched on.
It is based on what they call ``degree of confidence'', this parameter is calculated taking into account the number of vehicles at the intersection. 
The simple idea of adding a VANET in this kind of scenario raises an improvement.
This is analysed in \cite{desai2014instinctive}.
The problem with these systems is to forget the prioritization when emergency vehicles appear. 
They assume that the highest priority is given to them but none mechanism to do it is defined.

A complete system that allows us data extraction and processing of a VANET must be like the one proposed in \cite{abbas2011traffic}, in which we
have different controllers for the decision-making of changing traffic lights.
The problem is that they define the global system and the communications roughly, but they do not take into account the implementation of the system. 
They focus on the schema but not on the algorithms or on the controllers.

This paper is based on the work \cite{caballero18}, where a first approach to a collision avoidance system for traffic lights was proposed. The   system here presented includes a priority system for traffic lights that takes into account the operation in real time with the aim of improving the response times of emergency services. 
Besides, several improvements have been made in the simulation systems and in the reliability of the neural networks that has been used. 
In this way, better results have been obtained. 
In addition, new simulations and tests have been carried out to verify the right operation of the system. 
In this way, a more complete and robust system is here presented. 

In conclusion, the main goal of this work is to propose a cheap system to the management of dynamic traffic light systems.
The basic idea of this paper is to present a priority system for intersections controlled by VANET elements taking advantages of V2V and V2I communications, and in order to predict and/or detect red light running to warn nearby drivers and pedestrians. 
The objective is to try to avoid possible accidents and generate an efficient traffic light control.
To the best of our knowledge, there is no smartphone-based approach that automatically predicts and/or detects red light running, and reports the offense to nearby vehicles and pedestrians to prevent possible accidents without  any complex infrastructure.

It must be taken into account that the described application is proposed as a preliminary step of an internal implementation of the system in smart vehicles. In the proposal, it is assumed that most drivers currently use mobile applications with multiple services to drive, so that the proposed solution could be integrated into a route navigation application.

\section{Proposed System}

In this Section, the collision avoidance for red light system and the dynamic traffic light system are detailed.

\subsection{Collision Avoidance for Red Light System}

The operation of the proposed system is explained below.
Any vehicle with a smartphone on board  or pedestrian carrying one can take advantage of the system when approaching a traffic light with a sensor and communication module.
The application has two functions. 
First, it advises the driver to respect the traffic lights when they are in red, and sends warnings  both to  close vehicles and pedestrians  with its Wi-Fi radio interface when it predicts that the vehicle is about to run a red light, and to other vehicles and pedestrians  in the neighbourhood  through a central server if in fact it has run the red light. 
Second, it receives warning messages from nearby vehicles and central server, and presents visual and audio warnings to the driver/pedestrian. 
When the vehicle approaches the traffic light, the smartphone on board  receives  from the traffic light a beacon with its colour at that moment, and if it is in red, the driver is warned by the smartphone  with audio and visual signals. 
Then, if the driver's smartphone  predicts through the processed data that the vehicle is  not  stopping at the traffic light, it immediately sends a warning to close vehicles and pedestrians through Wi-Fi. Finally, if in fact he/she does not stop and run the red light, the driver's smartphone automatically detects it and sends a warning to a central server, which warns other  smartphones in the neighbourhood. In this way,  other drivers and pedestrians receive a warning about the dangerous situation through their smartphones.

\begin{figure}[!htb]
  \centering  
	\includegraphics[width=0.9\textwidth]{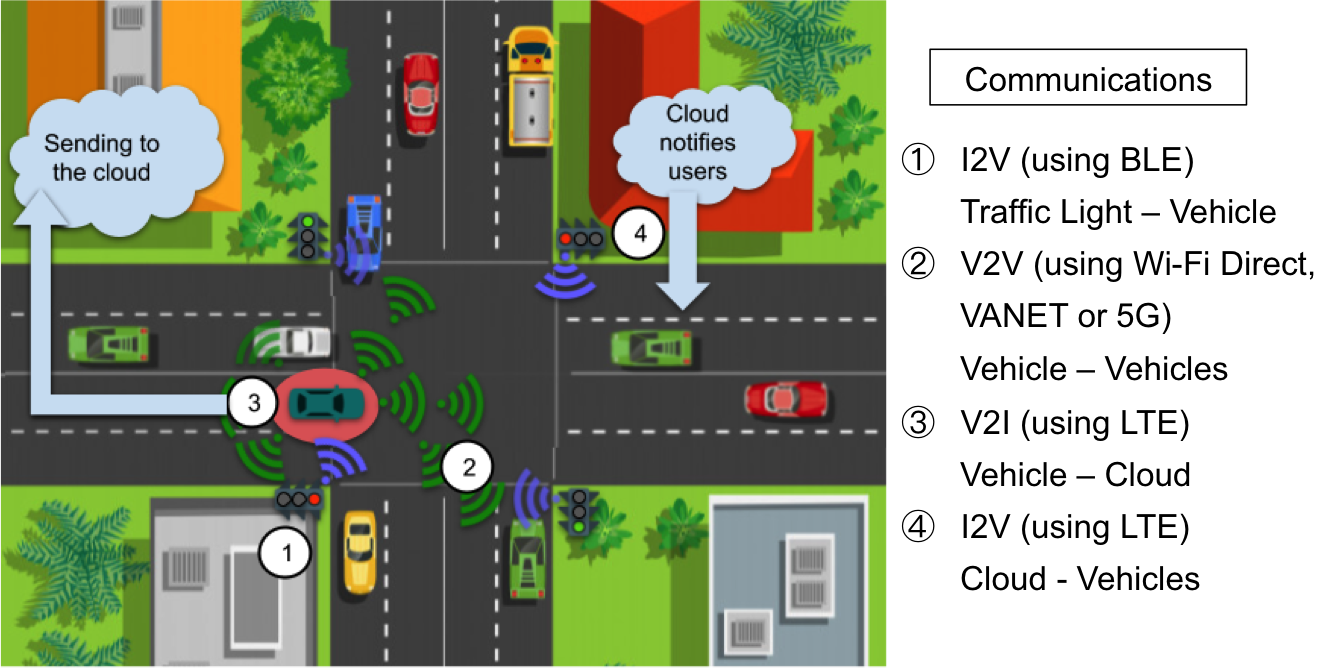}
  \caption{Communications within the System}
  \label{fig:cruce}
\end{figure}

The proposed system is based on  I2V, V2V, V2I and V2P communications (see  Figure \ref{fig:cruce}), and consists mainly of:
\begin{itemize}
\item Traffic lights: Equipped with light sensors and Bluetooth Low Energy (BLE) modules  to communicate  with nearby smartphones.
\item Smartphones: Located inside vehicles or carried by pedestrians,  to predict and detect red light running and to send warnings to other smartphones and to the cloud server.
\item Cloud servers: Responsible for receiving    red light running warnings and alerting neighbourhood smartphones   about the danger.
\end{itemize}

\begin{figure}[!htb]
  \centering  
	\includegraphics[width=0.8\textwidth]{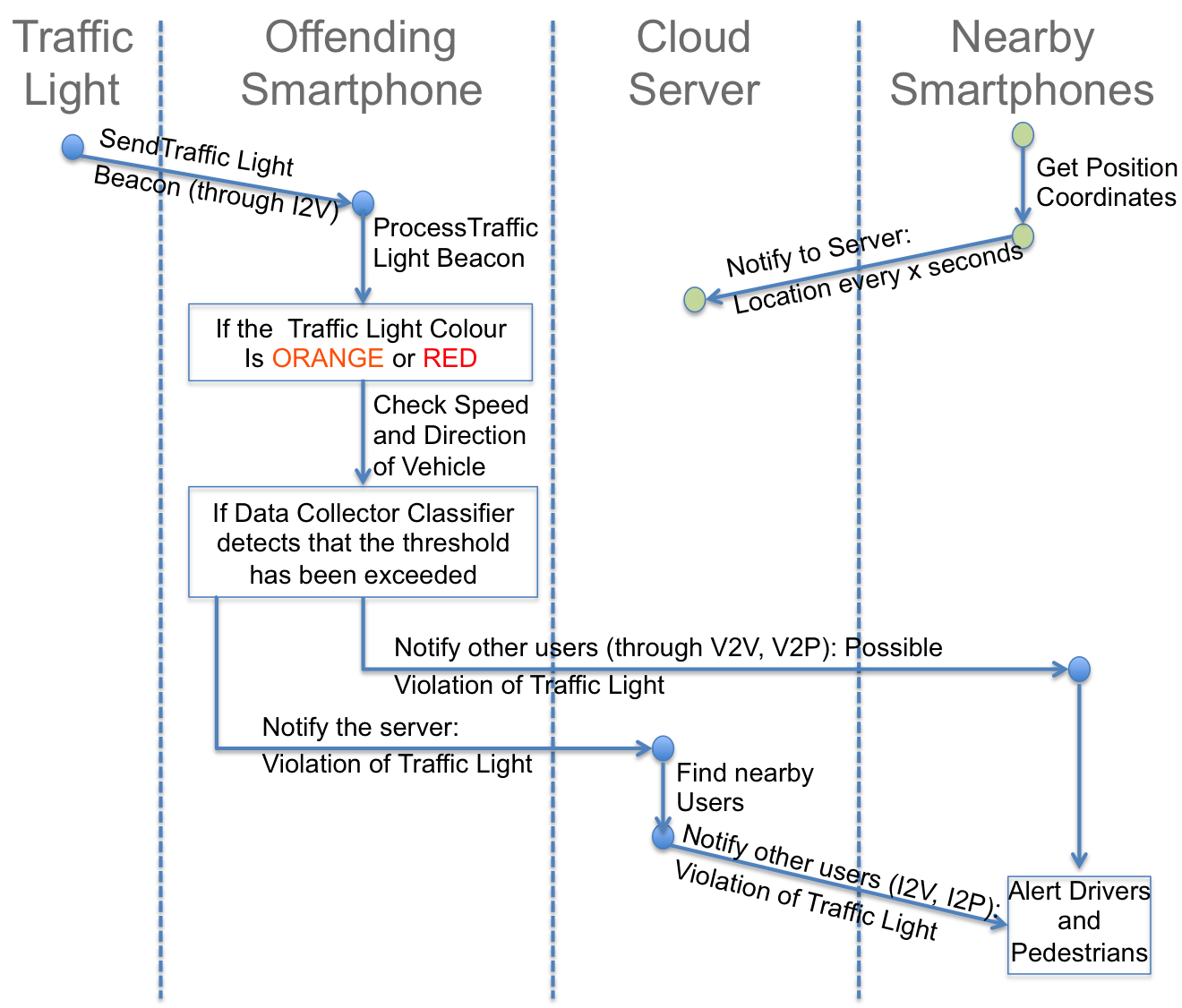}
  \caption{Overview of the System Operation}
  \label{fig:alg}
\end{figure}

Thus, the implemented system uses sensors, smartphones and cloud servers to automatically predict, detect and  report that a driver has failed to respect a traffic light. Figure \ref{fig:alg} shows an overview of the system operation. 

The proposed scheme requires that the traffic light   continuously send beacons with its colour. The beacon advertising interval of used can be set between 20 ms and 10.24 s, but in general it is recommended that it be at least 100 ms. In the implementation, the traffic lights send beacons with a frequency of 4 times per second, so the advertising interval is 250 ms. If it is orange or red, the smartphone aboard a vehicle approaching it, checks its speed and direction to verify that such a traffic light affects it. In that case, it  processes  collected data to predict a possible  red light running or to detect a red light running.

In this latter case, the cloud server receives a warning  message from the offending smartphone, finds nearby smartphones of vehicles and pedestrians  and sends them a push notification  to alert them of the danger.

Thus, apart from smartphones, the main technology tools  used in the proposed system are:
\begin{itemize}
\item Light sensors, to provide real-time information   about the  colour of the traffic light. 
\item BLE modules, to allow  continuous transmission   of   traffic light status to nearby vehicles, as  beacon notifications.
\item Arduino-compatible  devices, to allow  sending  data through  BLE.
\end{itemize}

\subsection{Dynamic Traffic Light System}

\begin{figure}[!h]
	\centering
		\includegraphics[width=0.5\textwidth]{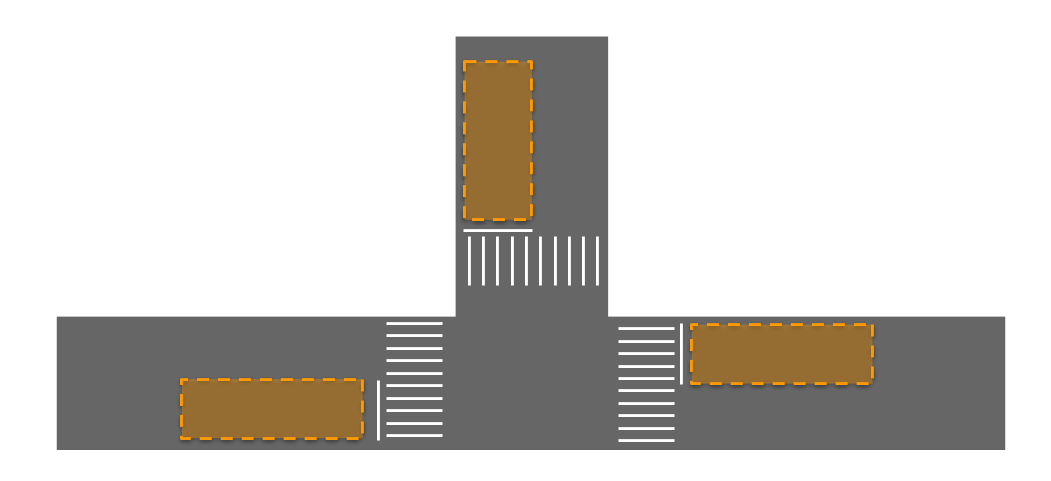}
		\caption{Studied areas for three branches.}
	\label{fig:areas_cruce}
\end{figure}

The basic approach of this prioritization system is based on changing traffic lights at intersection, according to the needs and characteristics of traffic in each moment.
To achieve this, the top priority in the system is assigned to emergency vehicles.
If emergency vehicles are not present at branches, priorities will be allocated based on the algorithm here proposed.
This algorithm analyses the characteristics of each branch and takes into account the amount of items affecting them.
Before explaining the algorithm, it should be considered that each intersection has a different structure. 
The corresponding area should be analyzed in order to calculate weights (see Figure~\ref{fig:areas_cruce}). 
In this example, an intersection of three branches with the same size area for each branch is represented. 

These areas must be predefined before implementing this system because they will be responsible for the operation. 
Additionally, in these areas the maximum numbers of vehicles that can be found \textit{(${N_{1}}$, ${N_{2}}$,..., ${N_{n}}$)} should be determine since this is a fundamental parameter for the decision system. Then, the maximum weight of each branch will be calculated.
The proposed algorithm tries to assess the weight of each area in
the branches in real time, considering the following parameters: \textit{``b''}
Number of branches, \textit{``n''} Number of vehicles, \textit{``e''} Number of
emergency vehicles and \textit{``p''} Branch priority (\textit{1} for a
priority road, \textit{0} for a secondary road). When the studied
branches have the same priority, it is not necessary to consider the
parameter \textit{``p''}.
After the analysis of these parameters, each of the involved
areas  has  a  weight   (${W_{1}}$,...,${W_{T}}$), which will correspond to the sum of three parameters:
${W_{n}}$ Normal vehicles weight, ${W_{e}}$ Emergency vehicles weight,
${W_{p}}$ Branch priority weight. That is, ${W_{T}}$ will be:
\begin{equation}
W_{T} = W_{n} + W_{e} + W_{p}
\end{equation}

The weight of the vehicles in our area ${W_{j}}$ will be simply the
number of vehicles n that are in it.
In the case of the weight of emergency vehicle's (${W_{e}}$), the number
of emergency vehicles (\textit{e}) will be considered. To ensure that it has
always the highest priority, \textit{e} will be multiplied by the maximum
number of vehicles that may be there (\textit{N}). Thus, whenever there is
an emergency vehicle, this branch will have a higher priority than
any other branch. When different sizes are presented, the higher N
will be chosen.

\begin{figure}[!h]
	\centering
	    \subfigure[Branch $i$ weight]{\includegraphics[width=0.45\textwidth]{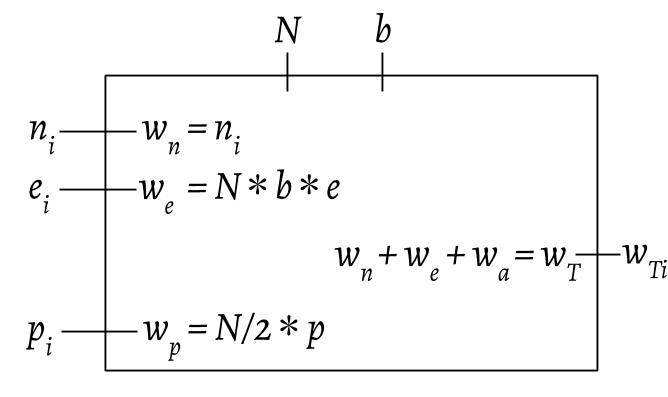}}
	    \subfigure[Priority selection.]{\includegraphics[width=0.45\textwidth]{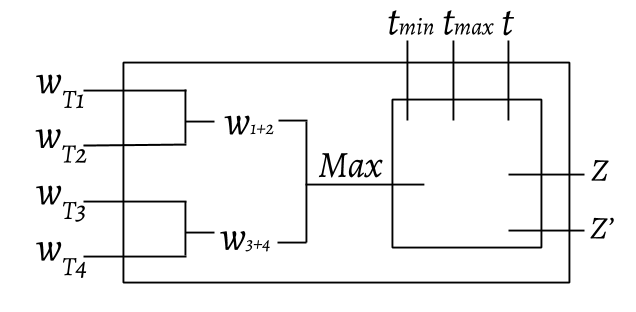}}
    \caption{Traffic Light Weights scheme}
	\label{fig:algoritmo_paso1}
\end{figure}

Finally, to calculate the weight of the road priority, ${W_{p}}$, the half of
the maximum weight will be added or not depending on $p$. 
As aforementioned, the value of $p$ is related to the priority of the branch. If there is a priority road, it has the value of $p$ set to \textit{p = 1}, so its ${W_{p}}$ is \textit{N/2} higher than that of non-priority roads, (where \textit{p = 0}).
In this way, if the road has not priority nothing is added, but if the road is the main its weight always will be \textit{N/2} higher than any other road. 
Finally, the weight of each area is calculated as it is showed in Figure~\ref{fig:algoritmo_paso1} (a).

When the weight of each branch is calculated, the direction is
taken into account, and the final weight is calculated. The branch
with the maximum weight is going to have priority (see Figure~\ref{fig:algoritmo_paso1} (b)).

There are some parameters that have not been discussed so far: 
\textit{``tmin''} minimum time that a traffic light should be green,
\textit{``tmax''} maximum time that a traffic light should be green and 
\textit{``t''} time that traffic light maintains that state. 

Thanks to the use of $tmin$ and $tmax$, the system can control the time to allow cars to pass comfortably. That is, on one hand, if the range between $tmin$ and $tmax$ is too short, cars may not have enough time to move on. On the other hand, if this range is too long, the system could make the vehicles wait forever.

\begin{figure}[!h]
	\centering
		\includegraphics[width=0.8\textwidth]{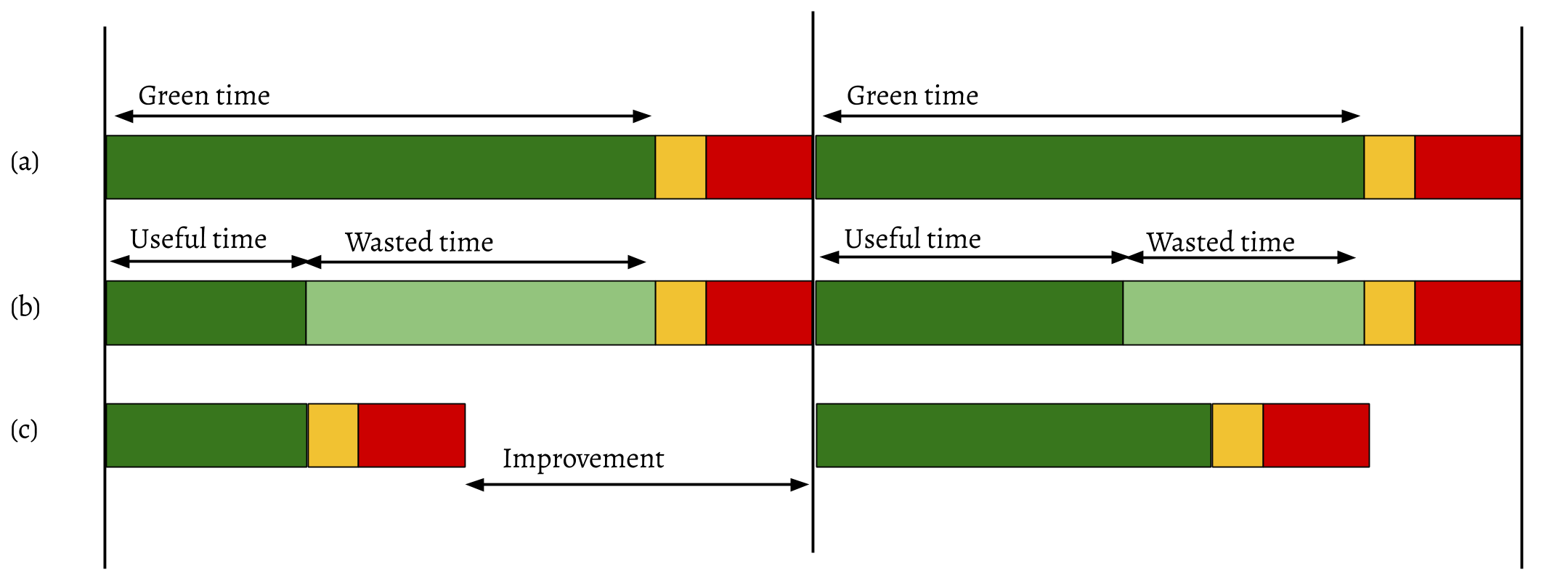}
		\caption{Traffic light comparison.}
	\label{fig:traffic_light}
\end{figure}

Once these times are known, the time of the proposed system may
be calculated. In traditional traffic lights, there is fixed time
duration in each state, as shown in Figure~\ref{fig:traffic_light}(a). However, it is not
always used for the passage of vehicles; there is usually wasted
time where vehicles do not pass, as shown in Figure~\ref{fig:traffic_light}(b). With
this system, it can be detected when the crossroad is not at its
peak performance and make a change to allow other routes to
pass. Thus waiting times of the vehicles with the system proposed
will be as far the current time-out with traditional system Figure~\ref{fig:traffic_light}(c).

\begin{figure}[!h]
	\centering
		\includegraphics[width=0.5\textwidth]{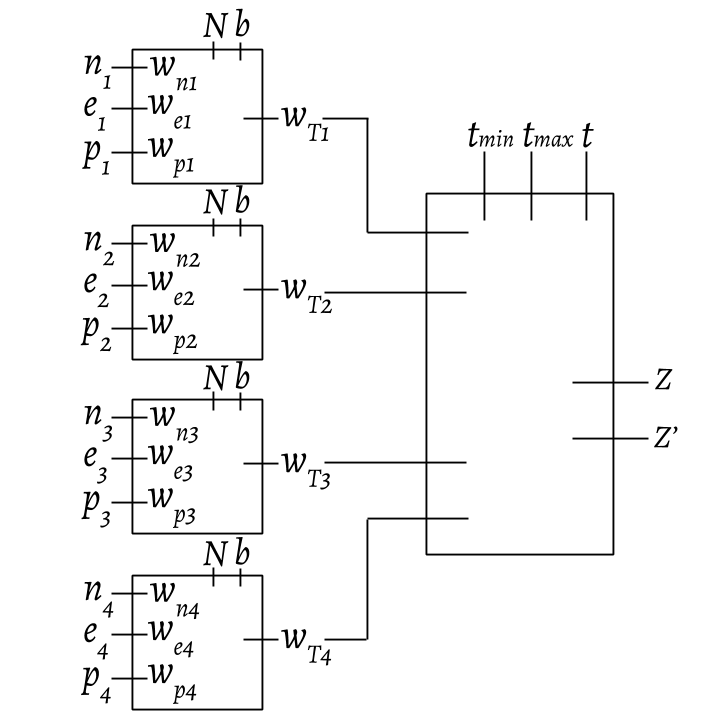}
		\caption{Global algorithm.}
	\label{fig:algoritmo_global}
\end{figure}

When dealing with waiting time for an emergency vehicle, it must
be minimized. The overall algorithm of the system is in Figure~\ref{fig:algoritmo_global}.

\section{User Application}

The proposal is based on the use of a mobile application to  read the BLE beacon emitted by the traffic light,   process the  collected data,  and send a warning to other smartphones and/or cloud server.
In order to monitor the GPS position  of all user smartphones, receive V2I notifications after a red light running, and send  I2V warnings to    neighbours of the offender, the use of a cloud server  is proposed. 

When the  beacon received by the user smartphone corresponds to an orange or red traffic light, depending on whether the distance to the traffic light is greater than zero or not, various collected data such as   position, speed and acceleration of the vehicle allow the application to estimate through the classifier if the vehicle is about to run the red light or has actually run the red light. 
In the first case, a warning is directly broadcast  to close smartphones via Wi-Fi. In the second case, a notification is sent from  the offending smartphone  to the cloud server that controls and manages those events. 
This  cloud server is responsible for searching its database to find nearby vehicles. This is possible because all vehicles periodically send  (for instance, every 5 seconds) their current positions to the server using the smartphone Internet connection. Once the server has located all user smartphones near the traffic light, a push notification is sent to all those smartphones. 

The estimation system included in the mobile application consists of two modules: data collector and classifier.
While the application is running, the data collector periodically obtains data from the GPS, the accelerometer and a map database. The post-processing of these data is done before the information becomes an input for the estimation model used by the classifier. For instance, the GPS location and map data are used to estimate  information such as distance to intersection, speed and acceleration.

\begin{figure}[!h]
	\centering
		\includegraphics[width=0.5\textwidth]{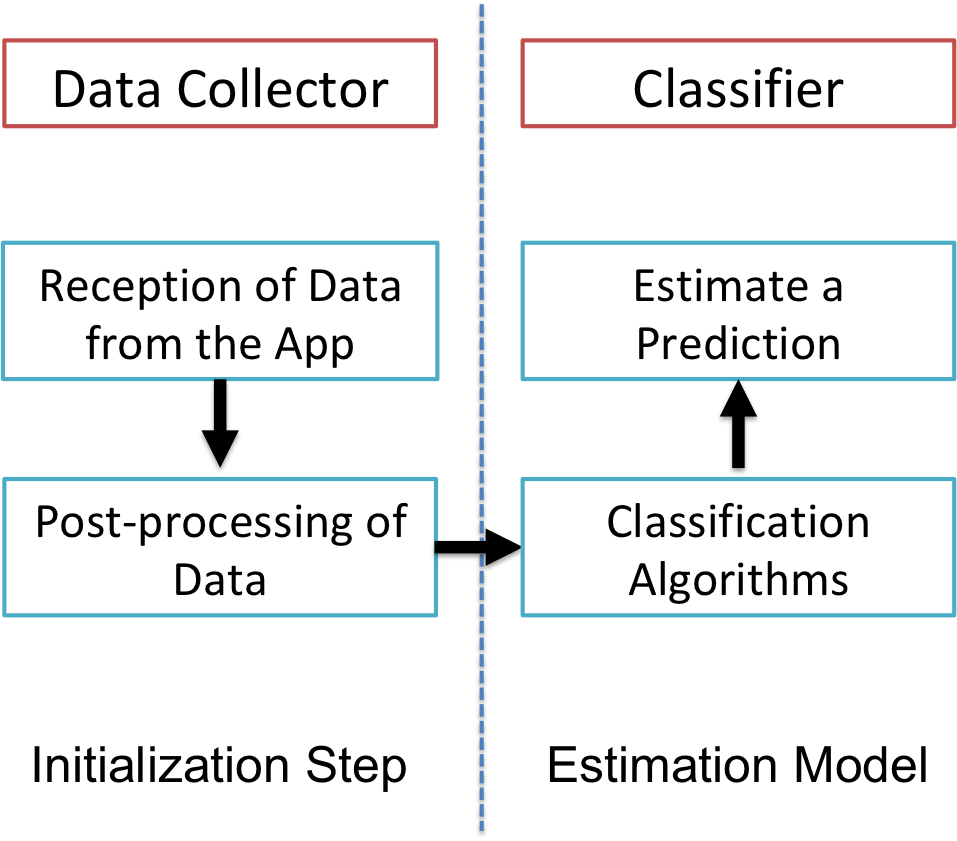}
		\caption{Modules}
	\label{fig:Modules}
\end{figure}

The  classifier takes the  data processed by  the data collector as input, and uses a set of classification algorithms to estimate a prediction  that a red light running is likely to occur, or a detection that  a red light running has actually occurred. Figure~\ref{fig:Modules} shows the two modules of the  mobile application and their relations.

There are some situations where the system assures that a red light running is going to happen. These situations are the ones where the distance to the traffic light is lower than the stop distance. The stop distance is the addition of the braking distance and the reaction distance. To calculate this distance, the formula shown in Equation (2), where the speed of the vehicle is taken into account, is used. The distance between the vehicle and the traffic light is easy to obtain due to the GPS position of the traffic Light is in a database that can be queried from the smartphone using the id of the beacon.
\begin{equation}
stop\_distance = speed / 10 * 3 + (1/2 (speed/10)^2)
\end{equation}

On the other hand, the estimation problem can be formulated as a supervised learning problem based on a binary function from labelled training data as can be seen below.

The used classification algorithm is based on the Support Vector Machine (SVM) model \cite{Chang}, which is a general-purpose machine learning algorithm. This algorithm is normally used in cases such as this one, when relatively small real-world traffic  data are available and no assumptions can be made about any particular data distribution. Moreover, its learning process also involves a selection of characteristics to be used, by automatically identifying the appropriate attributes to perform a satisfactory classification. The training data set consists of  several features used as input in vector form, and labelled with correct output values. Each function should produce as output  value  the estimation of a prediction or of a detection, when given a valid input.

The data must be collected at specific distances (every few meters) to the intersection, concretely the speed of the vehicle is measured at these distances. To obtain sufficient data to be used in the training phase of the SVM algorithm, the SUMO simulation tool has been used. Later, each speed vector obtained in the real-world use of the application is added to the dataset to improve the accuracy of the system.

In particular, the system used for estimation takes a set of speed data every 5 meters as the vehicle approaches to a  traffic light, before arriving at it. Moreover, each vector is labelled with 1 if bad driver behaviour has occurred, such as when a red traffic light running has happened, or with 0 if a good behaviour has been detected, such as when the driver has stopped at the red traffic light.  In order to make an estimation, the algorithm begins to collect the data when the vehicle is  at a configurable distance from the intersection, and stops when   it leaves the traffic light behind.  All data collected in that time period  are used as input to the SVM estimation model,  which   returns a confidence value representing the probability that the result is correct. Finally, the confidence value is compared to a configurable threshold in order to output the final estimation result: predicted/detected offense or not.

If  the confidence value of that estimation is greater than the configurable threshold,
then the  classifier reports that a red light running is likely to occur or that it has actually occurred. Specifically,   the prediction case is often erroneous because  most drivers do not decide whether to run the red light  until they get very close to it.

\section{Communications Security}

The proposal includes different types of communications, including I2V, V2V, V2P and V2I, which require different forms of protection.

First, the traffic light emits beacons that are transmitted through BLE so that the nearby vehicles receive and interpret them. Therefore, such I2V communication  must be protected against possible falsifications. This is achieved in the proposal by using  an authentication scheme with quick verification.

Second, the potential offending vehicle sends a warning directly through Wi-Fi via V2V/V2P communications to vehicles/pedestrians at the intersection, which also requires fast authentication. In this case anonymity is also required to encourage the use of the application since its purpose is not to punish the offender.  In case the system finally estimates that the traffic light has been skipped, the offending vehicle sends the LTE notification to the server so that it in turn alerts the surrounding vehicles and pedestrians. In this case, there are two types of communications that need to have the sender's authenticity and message integrity guaranteed. On the one hand, the V2I communication  requires signature with anonymity. On the other hand, the I2V/I2P communication  simply requires a digital signature.

The operation of the system is based on the fact that all vehicles periodically send their location to the server. This V2I communication needs to have its integrity and authenticity of sender protected by signature, although at the same time it is important for the protection of privacy, that the signature incorporates anonymity.

Whenever the integrity of a message and the authenticity of the sender have to be protected by the use of a digital signature scheme, this implies that the sender has to use  its private key during the digital signature process, and that the receiver must use the sender's public key to verify the digital signature of the message. Prior to this, the sender's public key must be  verified through its certificate, which is signed by a Certificate Authority (CA). All this means that signed messages involve overhead mainly due to the certificate and signature that must be attached to the message. In addition, the  verification of such signatures also leads to a high computational cost. Therefore, in   cases where sender authentication  is not required and message integrity  must be protected by  an efficient procedure, the proposal involves the use of a Message Authentication Code (MAC). A MAC is generally based on a cryptographic hash function and a secret key in order to allow verifiers who share the secret key with the sender to detect any changes to the message content.

As for the I2V communication from traffic light to nearby vehicles, since it is based on BLE technology, it is known that, according to its technical specification, it uses  128-bit AES with Counter Mode CBC-MAC. Apart from that, in the proposal, 
 beacons are authenticated using a MAC with a secret  key shared among users in the neighbourhood. To do this,  when the server  detects that a vehicle is in the vicinity of a traffic light, it send the shared secret key corresponding to that traffic light if it does not already have it.

The current IEEE Trial-Use standard  for VANET security \cite{IEEE} recommends some cryptosystems. According to it, to authenticate the sender of a message  and  ensure the integrity of the message, OBUs or RSUs must sign each message  with their private keys before sending it. The proposed cryptographic scheme is the Elliptic Curve Digital Signature Algorithm (ECDSA) \cite{ecdsa}, which is a variant of the Digital Signature Algorithm (DSA)  using elliptic curve cryptography.
One of the problems of this standard is the cryptographic overhead caused because the certificate and the signature  take up a significant portion of the total packet size.
In addition,  cryptographic operations also lead to high computational costs for receivers when verifying these messages. If a vehicle sends a message within a time interval of 100-300 m  to 50-200 vehicles within its communication range, each receiver needs to verify around 500-2000 messages/s, including the signed public key certificates contained in them.  Therefore,  verification algorithms are required to be  fast enough to allow   incoming messages to be processed. Unfortunately, PKI-based schemes are not suitable for this strict time requirement, so it is proposed to use  MAC as a solution to protect message integrity and source authentication,  guaranteeing low communication overhead and fast verification. In addition, to promote the use of the proposed scheme, the solution should protect privacy and/or anonymity. Therefore, vehicle identities must be hidden during the authentication process to protect private information from senders, such as the identity of the driver and other personal information. On the other hand, the authorities should be able to trace the sender of a message by revealing its identity for example in case of an accident.

In order to protect user privacy, the concept of $k$-anonymity, where $k$ entities are indistinguishable, \cite{sweeney} is used to mix $k$ vehicles.
In the proposal, the server assigns a common pseudo ID to $k$ vehicles, where the $k$ vehicles (as a group) will take the same pseudo ID so that  an adversary cannot track a specific vehicle through its pseudo ID.

In this scenario, there is no possibility of homogeneity or similarity attacks \cite{Machana2006} because it is not necessary to share data among vehicles. The required data is about the vehicle or the driver, and for this proposal, not many details are required, only those necessary to broadcast the information.
If the authority needs to identify a driver, it uses the pseudo changing IDs in the shared packets. The authority stores the real identification, but no sensible data about vehicles or drivers are sent. Thus, there is no possibility of homogeneity attacks with this data.

On the one hand, in the MAC with $k$-anonymity used in the proposal, each pseudo ID is linked to $k$  secret keys. Thus, if the local server distributes secret keys shared in groups formed by neighbours, all members of each group can directly verify the MAC issued by another member of the group. In the remaining cases, it is considered the possibility of sending the received warning  to the server, so that it can verify its authenticity through the attached MAC, and in case of failed verification notify the nearby vehicles.

On the other hand, the  digital signature with  $k$-anonymity used in the proposal is based on the application of the ideas in \cite{caballero16}, according to which  every user is randomly associated with a group that shares cryptographic material such as a pair of private-public keys and a group certificate for these data to be used to sign. In this way, users do not reveal their particular identities, but only their group identifier. In this case, the proposal uses a group signature, which is a digital signature based on a group public key and individual private keys defined by a group manager so that it can be used to sign on behalf of the group and any signature can be verified with the group public key, but only the group manager can identify which member issued a given signature. In our case, the group is the group of vehicles, and the  group manager is an official entity related to road traffic. An additional advantage  of using group signature is the  efficiency of verification because verification of group signatures is done with respect to the public key of the whole group.

Everybody can check the trust of the signature by using the public key of the group, however,  there is an official member related to the road traffic that can verify the identity of a particular member.

There are different algorithms \cite{Aya2014} to retrieve the user from the k-anonymized group. There is no best anonymization algorithm for all scenarios, but the best performing algorithm in a given situation is influenced by multiple factors.
This work is related with VANETs, and part of  the k-anonymity  scheme proposed here is presented in \cite{caballero16}. In that work, the vehicles joined groups to guarantee that the tracking of a vehicle is not possible. The difference of the scheme here used is that in this case, the authority can recover the identity of a user in case of a serious incident.  All the information related to the processing time is presented in the aforementioned work.

The proposed $k$-anonymity system does not always prevent the disclosure of confidential attributes, such as if the vehicle is about to run a red light or has already run it. This fact allows that the privacy of drivers to be maintained  through reversible anonymity so that only fraudulent drivers who cause accidents lose their privacy. This is possible by making that the intersection between each group  sharing pseudo IDs and each group  sharing secret keys or pairs of   private-public keys contains only one element. On the one hand, when using MAC, the server can  identify a vehicle by finding the secret key paired with the pseudo ID. On the other hand, when  digital signature is used, the server can  identify a vehicle by finding the pair of private-public keys paired with the pseudo ID.

When an offence is automatically broadcasted, not only authenticity of sender and integrity of data are guaranteed, but also freshness of data. Thus, in order to guarantee freshness, a timestamp is added to  signatures.

\section{Implementation Analysis}

\subsection{Collision Avoidance System} 

A beta version has been implemented with some parts of the proposal.
It includes a sensor  connected to the traffic light to capture  its colour, and an Arduino  device to transmit the light status through a beacon advertisement.
On the one hand, the traffic light sensors chosen for  the implementation are small electronic modules that can be incorporated at any type of traffic light. On the other hand,   RFduino \cite{rfduino}, which is an Arduino shrunk to the size of a fingertip and made wireless connection, is used. In particular, the 2216 model, with a Dual AAA Battery Shield, was chosen for the implementation. The shield has a step-up switching regulator that allows the batteries to be drained down to low voltages while still providing a stable 3.3V to the RFduino.
The BLE module used for the RFduino is the RFD22102 RFduino DIP. 

The format of  BLE messages includes a 1-byte preamble, 4-byte access codes correlated with the used RF channel number, and a Packet Data Unit (PDU) that can be from 2 to 39 bytes and 3 bytes of CRC. Thus, the shortest packet would have 10 bytes and the longest packet would have 47 bytes. Transmission times for these packages range from 8 microseconds to the smallest packet up to 300 milliseconds for the largest. The PDU for the advertising channel consists of the 16-bit PDU header and, depending on the type of advertising, the device address and up to 31 bytes of information. In addition, the active scanner can request up to 31 bytes of additional information from the advertiser if the advertising mode allows such an operation. This means that a considerable amount of data can be received from the advertising device even without establishing a connection. Advertising intervals can be set in a range from 20 m to 10 s, which specifies the interval between consecutive advertising packets. Thus, in the beta implementation we decided to use the 
ISO/IEC 9796-2 \cite{isoiec} scheme, which is a standard signature scheme widely used in the smart card industry for public key certificates and message authentication based on SHA-1 hash and RSA  digital signature, because its length is only 22 bytes, so it fulfils the storage requirements of BLE beacons. In the beta implementation, we also decided to consider that all the traffic lights share the same generic certificate to sign beacons, given by the CA of the department for  road traffic.

\begin{figure}[htb]
  \centering  
	\includegraphics[width=0.9\textwidth]{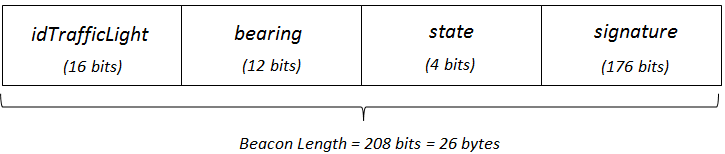}
  \caption{Format of  Beacon Sent by  Traffic Light}
  \label{fig:beacon}
\end{figure}

Thus, the beacon is formatted as shown in Figure \ref{fig:beacon}, where:
\begin{itemize}
\item {idTrafficLight}: Unique identifier for the traffic light.
\item {bearing}: Compass direction used to describe the direction of the traffic light (represented in degrees (0-360)).
\item {state}: State of the traffic light (green, red, etc.)
\item {signature}: Digital signature of the message.
\end{itemize}

The beacon is received by any smartphone on board an approaching vehicle at a maximum distance of 100 meters. Such a smartphone is responsible for processing information and self-reporting  any warning regarding potential red light running.  

\begin{figure*}[htb]
  \centering  
	\includegraphics[width=0.9\textwidth]{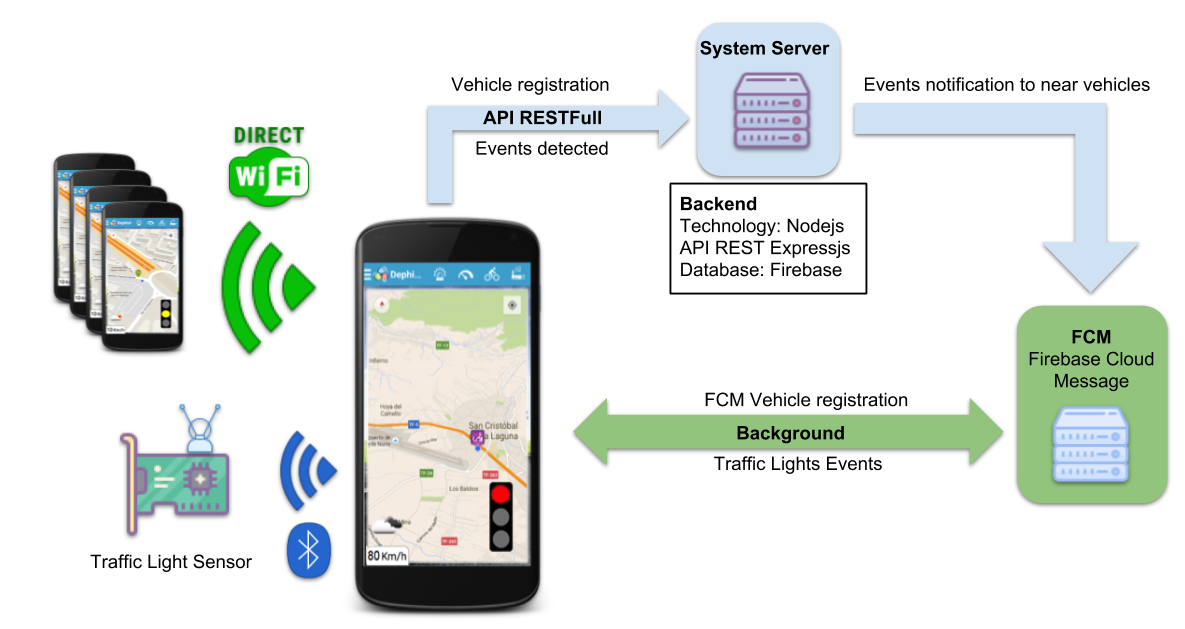}
  \caption{Use Flow and Technologies used in the System}
  \label{fig:tech}
\end{figure*}

The  system technologies used in the implementation are shown in Figure \ref{fig:tech}.

With respect to  the classifier implemented as part of the  Android application,  a distributed software library of the SVM model for   Java has been used.
Once  the vectors are obtained, and the training  and test samples are classified so that true positive, false positive and false negatives can be computed, those values give the accuracy of the predictions of the SVM model.

When the  classifier concludes a prediction or detection of a red light running, the warning message transmitter immediately begins to mount  a packet  whose payload  includes the   vehicle's GPS coordinates (4 bytes),  its speed   (2 bytes), its direction (2 bytes), and its identification  (plate number) (7 bytes).
The generated packet is then sent signed and encrypted either  to nearby smartphones or to a server.

Smartphones   equipped with an IEEE
802.11b/g/n Wi-Fi radio are used  to transmit and receive warning messages. Therefore,  to allow
Direct Wi-Fi communications between  smartphones, it is necessary to configure  the radio in
the ad-hoc mode.

The  cloud server used in the beta implementation is a full-stack JavaScript implementation. As cross-platform runtime environment for server-side and networking applications, Node.js is used. In order to connect the mobile application with the server through REST Web Services, the implementation uses Express.js, and to store the vast amount of data on the server, a NoSQL database called Firebase is used.

The implemented system applies different processes to send  distinct data packets. In Table \ref{tab:tam}, the size of the different data packets used in the proposed system is shown, where  packets are sent from traffic light to smartphone via BLE and from smartphone to smartphone/server via Wi-Fi/LTE.

\begin{table}[!htb]
	\begin{center}
	\begin{tabular}{| l | c |}
	\hline
	\textbf{Packet} & \textbf{Size} \\
    \hline
  	Beacon from Traffic Light to Smartphone & 208 \\
    \hline
	Event from Smartphone to Smartphone/Server & 248 \\
	\hline
	Notification from Cloud Server to Smartphone & 272 \\
    \hline
	\end{tabular}
	\end{center}
	\caption{Size of Sent Packets in Bits}
	\label{tab:tam}  
\end{table}

Each time a warning message is received by a smartphone, it immediately displays a large warning message on the screen and begins to sound a warning (see Figure \ref{fig:tech}).


\subsection{Traffic light system simulations}

Simulations have been performed by using SUMO (Simulation of Urban Mobility). The objective was to evaluate improvements over the current traffic light system taking into account emergency vehicles.

SUMO is an open source tool \cite{SUMO} that allows modelling traffic systems with all parameters, such as road vehicles, pedestrians, public transport, etc. 
It allows the simulation of road traffic to road networks.
Specifically, in this work, the logic of the system was developed in Python, thanks to the use of Traffic Control Interface (TraCI), which is a useful and popular Application Programming Interface (API) developed to establish communication between Python and SUMO.

\begin{figure}[!h]
	\centering
		\includegraphics[width=0.9\textwidth]{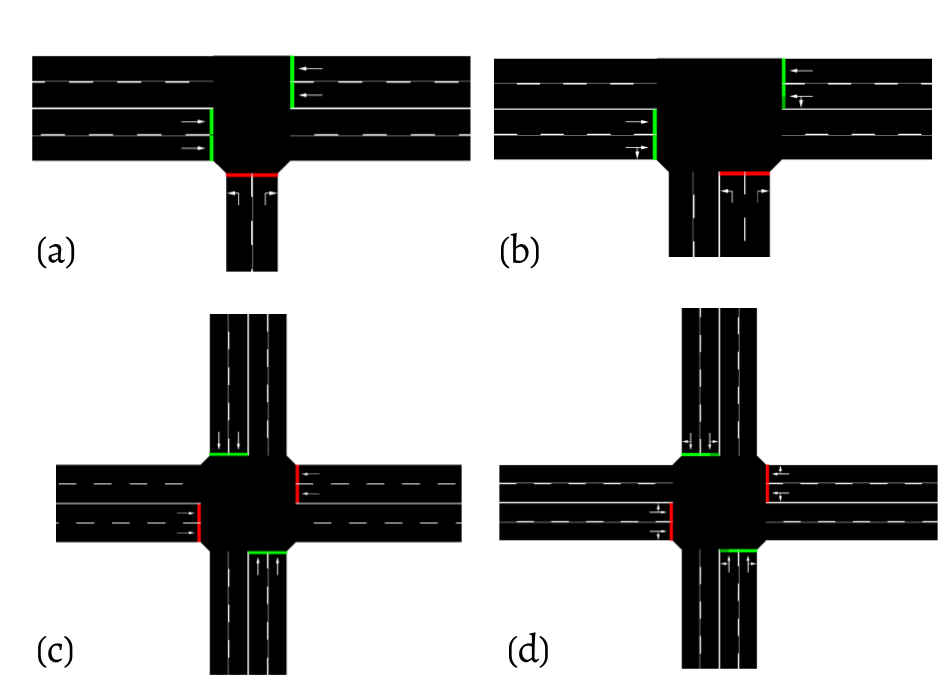}
		\caption{ SUMO scenarios.}
	\label{fig:sumo}
\end{figure}

Different scenarios have been generated with normal and emergency vehicles.
The generated simulations focus on the demonstration that a static traffic light system produces more wasted time than the proposed system. Taking into account the state of the road in real time can significantly improve the "waiting time" at crossroads of both normal and emergency vehicles (ambulances, police cars, fire trucks, etc.).
The tested scenarios are in Figure~\ref{fig:sumo}. 
There are two crossroads with three branches; one with reduced mobility (Figure~\ref{fig:sumo}(a)) and another more dynamic (see Figure~\ref{fig:sumo}(b)).

The other two scenarios generated are crossroads with four branches, one where only roads cross (Figure~\ref{fig:sumo}(c)) and another in which roads moves are linked (Figure~\ref{fig:sumo}(d)).

\begin{figure}[!h]
	\centering
		\includegraphics[width=1\textwidth]{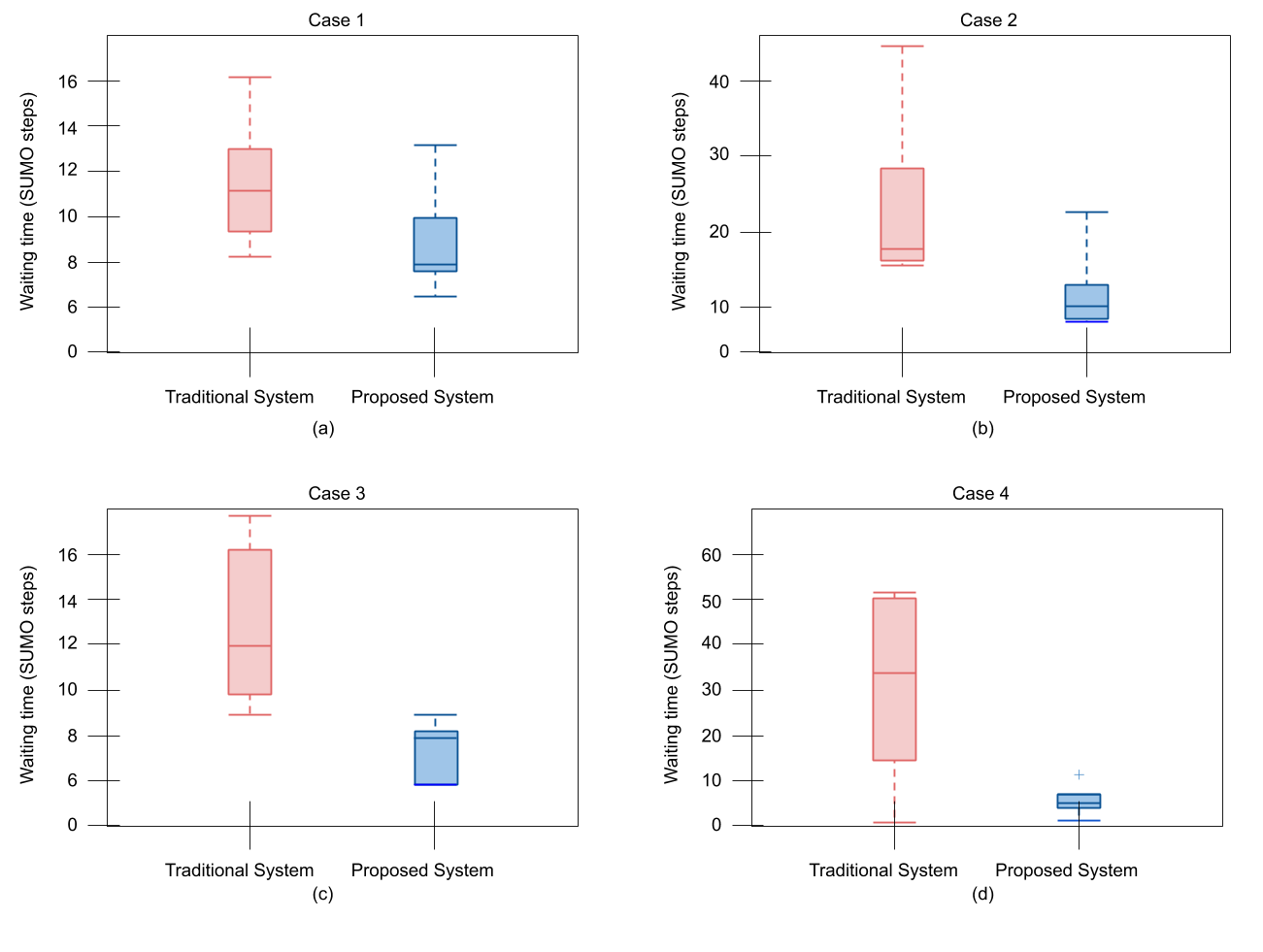}
		\caption{Systems comparison.}
	\label{fig:solutions}
\end{figure}

Each scenario was tested with the same features for both systems.
The behaviour of emergency vehicles on different routes was evaluated. 
Waiting times of emergency vehicles where collected
(results in Figure~\ref{fig:solutions}).
As aforementioned , the generated simulations focus on the demonstration that a "traditional system" with static traffic lights generates more wasted time than with the proposed system because knowing the state of the road in real time can significantly improve the "waiting time" at crossroads.

It must be noted that the emergency vehicles waiting times are measured with respect to SUMO, which allows us to measure the number of iterations performed. 
Each figure corresponds to behaviours obtained in each of the above scenarios. 
Case 1 and Case 2 scenarios correspond to the intersection of three branches
(see Figure~\ref{fig:solutions}(a) and Figure~\ref{fig:solutions}(b)). 
As more moves are allowed waiting times increase. 
In the traditional system there is a significant increase on this parameter. 
In both figures we can see that there is a significant improvement in waiting times when
using the proposed system. 
Case 3 and Case 4 correspond to scenarios with an intersection of four branches (see Figure~\ref{fig:solutions}(c) and Figure~\ref{fig:solutions}(d)). 
In these cases the time differences are greater than in previous cases.

\subsection{BLE and Wi-Fi Transmission Time analysis}

In real time systems like the presented in this paper the time that the signal is received by the users is vital. This time is known as transmission time, and depends on the distance from the user smartphone to the transmitter, the BLE beacon in the traffic light and the offender smartphone in the presented proposal. To measure this times a simulation analysis has been implemented. 

\begin{figure}[!h]
	\centering
		\includegraphics[width=1\textwidth]{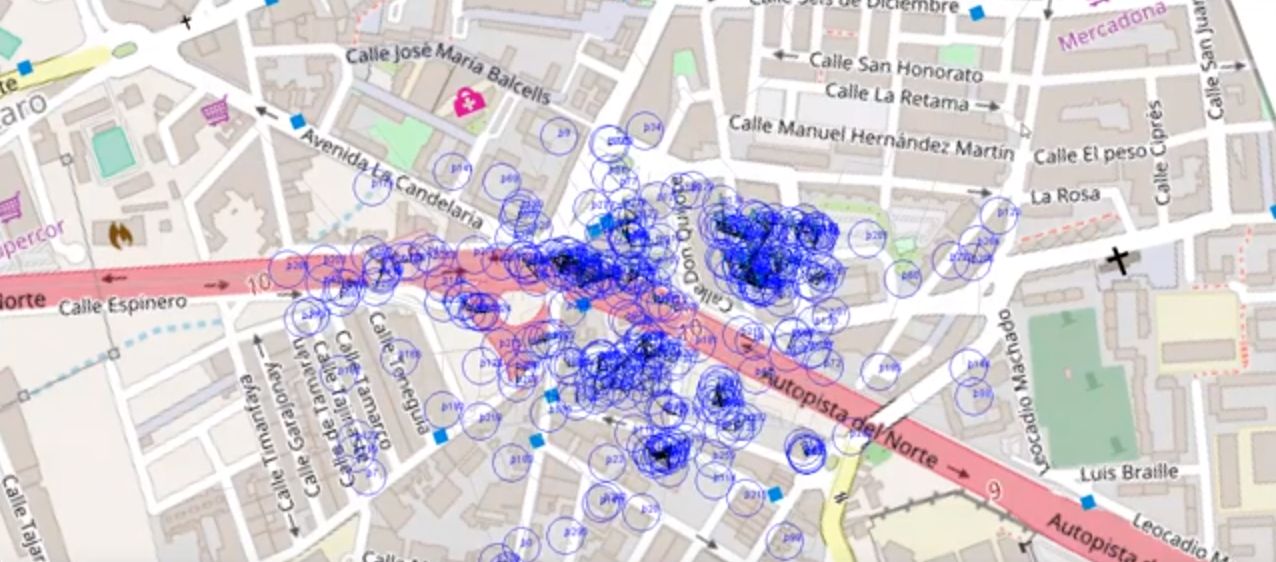}
		\caption{BLE beacon transmission time simulation}
	\label{fig:btsimu}
\end{figure}

To simulate this behaviour, the Opportunistic Network Environment simulator (ONE) tool has been selected. This tool allows to select different communication technologies, interfaces, nodes behavior, nodes speed, nodes deployment, simulation time, etc. Using these settings, a scenario where multiple devices send and receive data has been created. Around the simulated area, the number of 300 people affected by different red light running has been selected. These affected people can walk slowly, concretely between 0.05 m/s and 0,1 m/s varying randomly. With this walk speeds, the affected people can move between 200 and 400 meters in one hour, that is the time of the simulated event. The different nodes send new messages in a period of time that vary between 10 and 60 seconds to fit with this situation. In this analysis, the same scenario using the different available technologies (BLE and Wi-Fi) has been tested.
Moreover, the simulations have been performed 10 times for each combination of technologies to obtain
enough data. An example of the simulation scenario can be observed in Figure \ref{fig:btsimu}.

Using the data obtained in the simulations, the average transmission time using the two technologies can be observed in Table \ref{tab:time}.
\begin{table}[!htb]
	\begin{center}
	\begin{tabular}{| l | c |}
	\hline
	\textbf{Technology} & \textbf{Transmission Time} \\
    \hline
  	Bluetooth Low Energy & 232.7065 ms\\
    \hline
	Wi-Fi Direct & 569.3793 ms\\
	\hline
	
	\end{tabular}
	\end{center}
	\caption{Transmission Time of Wireless Technologies}
	\label{tab:time}  
\end{table}

This table shows how in average the BLE latency is more than 2 times less than the Wi-Fi. These measurements can be explained by two factors. The first of them is that in the case of Wi-Fi communications the two devices are in movement and in the case of the BLE beacon it is static in the traffic light. The second factor is that in the car to car communications via BLE the transmitter and receiver devices are inside the car where could be magnetic interferences. These two factors could explain the obtained results.

\section{Conclusions}
The systems proposed in this work adds a solution for two of the more common problems related with traffic lights, the red light running and the possibility of collision due to this, and the time optimization with traffic light in cross roads.
The collision avoidance  system proposed here allows automatic  detection of red light running and  warning to nearby vehicles and pedestrians  in real time in order to try to prevent possible accidents. The  described system  uses light sensors placed at normal traffic lights, which emit their current status as BLE beacon data. Each vehicle is paired with the driver's smartphone, which  is responsible for collecting and processing these  beacons together with other data related to speed and location in order to estimate through a machine learning algorithm whether the vehicle is about to run a red light or has already run it. 
On the one hand, if a  red light running is predicted,   the smartphone  of the offending driver directly  notifies it through Wi-Fi to other nearby smartphones of drivers or pedestrians. On the other hand, if a red light running is detected, smartphone  of the offending driver notifies it   to a local cloud server that alerts other vehicles and pedestrians in the area. 
Different cryptographic protocols are used to protect authenticity and integrity of messages sent from traffic lights, smartphones and servers, and privacy and anonymity to promote the use of the system.

The dynamic prioritization system at crossings based on weights was presented, paying special attention to emergency vehicles and its integration into a VANET. In the simulations, significant improvements in vehicle waiting time were obtained.
A beta version  with some parts of the proposal  has  been implemented   and the first results are promising. 

\section{Acknowledgments}
Research supported by TESIS2015010102, TESIS2015010106, RTI2018-097263-B-I00, C2017/3-9, IDI-20160465 and DIG02-INSITU.



\bibliographystyle{ACM-Reference-Format}

\end{document}